\documentstyle[11pt,newpasp,twoside,epsf]{article} 

\begin{document}
\title{Cluster Galaxy Populations from $z=0.33-0.83$:  Characterizing
Evolution with CL1358+62}

\author{Kim-Vy Tran and Garth D. Illingworth}
\affil{Department of Astronomy and Astrophysics, University of
 California, Santa Cruz, CA 95064}
\author{Marijn Franx}
\affil{Leiden Observatory, P.O. Box 9513, NL-2300 RA Leiden, Netherlands}

\begin{abstract}
We combine HST WFPC mosaics of clusters at multiple redshifts with
Keck LRIS spectroscopy to characterize different galaxy populations
and how they evolve.  By combining quantitative measurements of
structural parameters with velocity dispersions, line indices, and
morphological types, we attempt to establish links between
ellipticals, S0's, E+A's, and later type galaxies.  Here we focus on
galaxies in CL1358+62, a massive cluster at $z=0.33$.
\end{abstract}

\section{Introduction}

Using both high resolution imaging from HST WFPC2 and spectroscopy
from LRIS on Keck, we study the cluster galaxy populations in
CL1358+62 using 142 confirmed members.  The imaging provides colors,
morphological type, and quantitative structural parameters such as
bulge/total fraction, bulge and disk scale lengths, half-light radius,
galaxy asymmetry, and ellipticity.  The spectroscopy provides
membership, velocity dispersions ($\sim$masses), and line indices for
measuring recent star formation, e.g. H$\delta$ and [OII].  With our
complementary approach, we can place constraints on the evolution of
cluster early-type (E/S0) and post-starburst (E+A) galaxies in
CL1358+62.

\begin{figure}
\plottwo{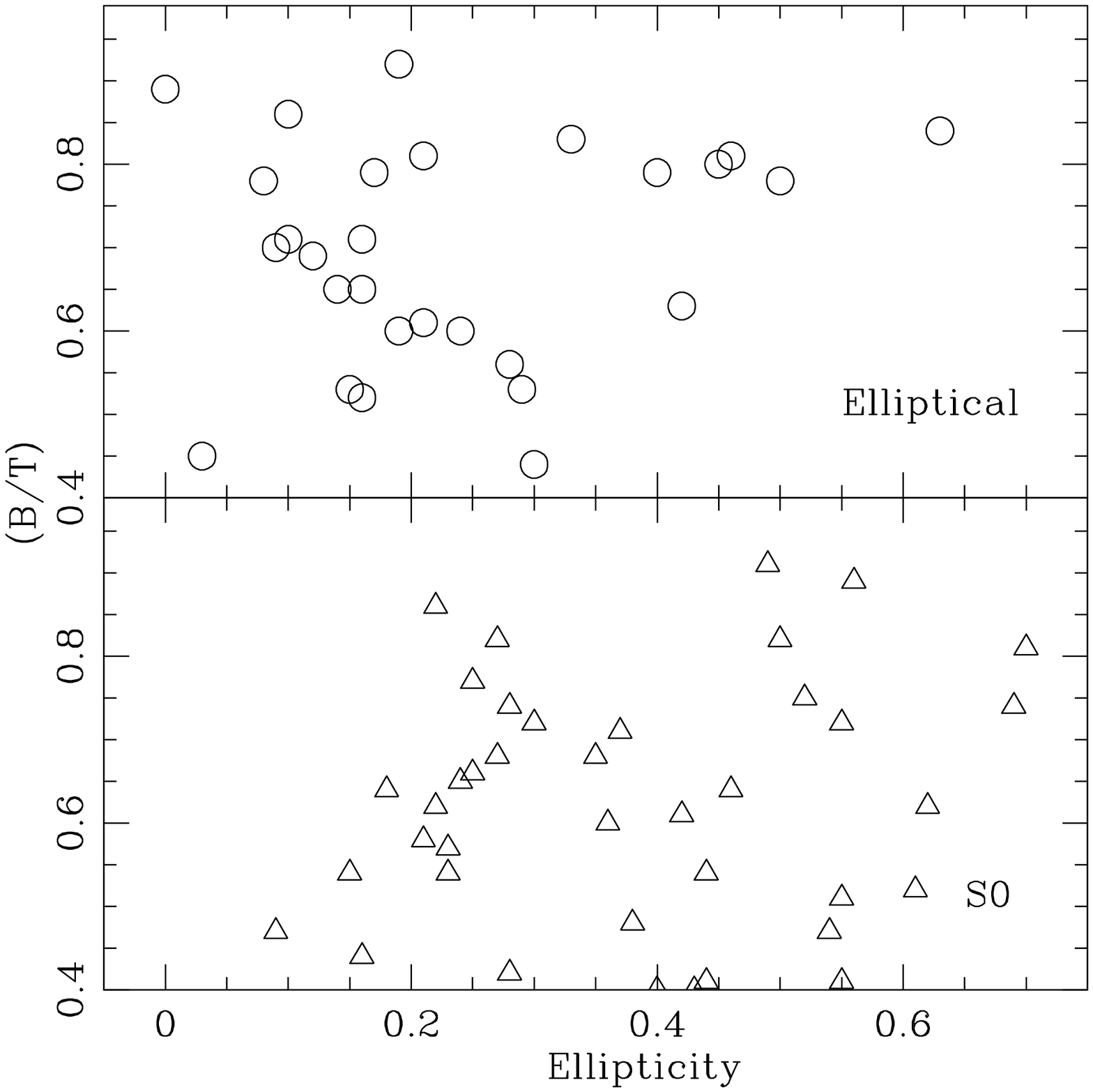}{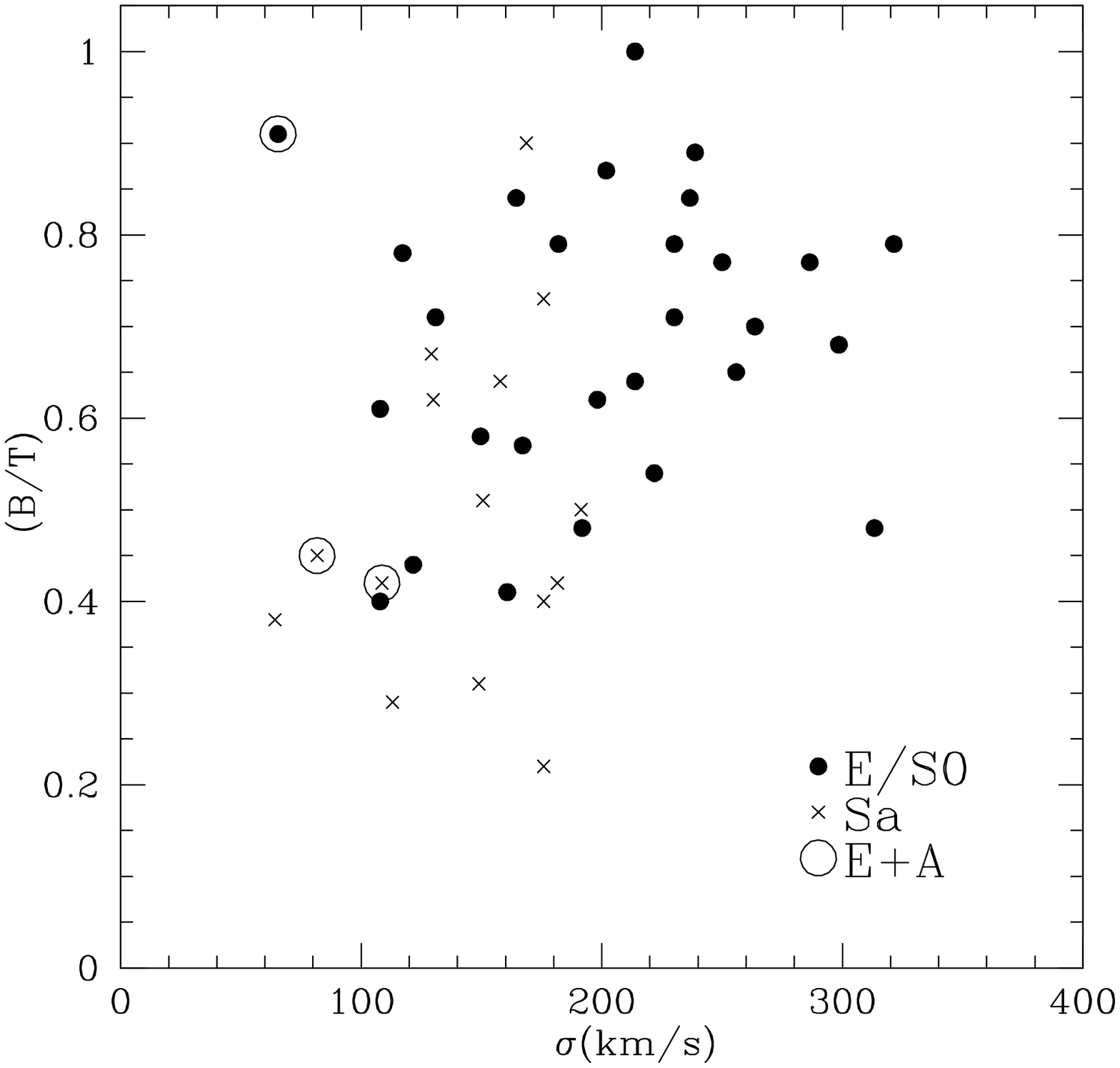}
\caption{{\it Left:} Bulge fraction vs. bulge ellipticity for
ellipticals and S0's; we show only the most bulge-dominated galaxies
($B/T\geq0.4$).  The $B/T$ distributions of E's and S0's are
indistinguishable while their ellipticity distributions do not share a
common parent population ($>99$\% confidence).  {\it Right:} Velocity
dispersion (Kelson et al. 2000) vs. $B/T$ for a subset of E/S0, Sa,
and E+A members.  As the E+A's in CL1358+62 do not share the same
velocity dispersion distribution as the E/S0 population, they cannot
evolve into massive E/S0 galaxies.}
\end{figure}

\section{Elliptical vs. S0:  A Matter of Viewing Angle?}

Jorgensen \& Franx (1994) suggest that elliptical and S0 galaxies
($M_B\geq-22$) are of the same galaxy class where E's are face-on
members and S0's the more edge-on ones, $i.e.$ these are the same
galaxies where viewing angle is the primary discriminator.  If the
CL1358+62 sample has ellipticals that are mainly round and S0's that
are elongated while both share a similar bulge fraction ($B/T$)
distribution, this would support a common parent population for both.
We show the ellipticity distributions of E's and S0's in Fig.~1
(left).  Results from the Kolmogorov-Smirnov tests of the $B/T$ and
ellipticity distributions for this sample supports a common parent
population for ellipticals and S0's.

\section{E+A Galaxies:  What Do They Become?}

E+A galaxies are still a mystery (Dressler et al. 1997).
Spectroscopically defined as galaxies that formed stars within the
recent past ($<1.5$ Gyr; Balogh et al. 1999), they may be the
transition between late type spirals and early type systems.  If E+A's
are a snapshot in the transformation of the former to the latter, they
are a fundamental phase in galaxy evolution.

To test if E+A's can evolve into the bright, massive early type
cluster galaxies observed nearby, we examine the bulge fraction, bulge
scale length, and velocity dispersion of E+A's in CL1358+62.  Their
scale lengths and velocity dispersions (Fig.~1, right) show a
population different from E/S0 galaxies.  The E+A's in CL1358+62 may
evolve into early type systems but they cannot become massive ones.

\end{document}